\input harvmac
\def\fig#1#2#3{
\par\begingroup\parindent=0pt\leftskip=1cm\rightskip=1cm\parindent=0pt
\baselineskip=11pt
\global\advance\figno by 1
\midinsert
\epsfxsize=#3
\centerline{\epsfbox{#2}}
\vskip 12pt
{\bf Fig. \the\figno:} #1\par
\endinsert\endgroup\par
}
\def\figlabel#1{\xdef#1{\the\figno}}
\def\encadremath#1{\vbox{\hrule\hbox{\vrule\kern8pt\vbox{\kern8pt
\hbox{$\displaystyle #1$}\kern8pt}
\kern8pt\vrule}\hrule}}

\overfullrule=0pt

\Title{MIT-CTP-2545, TIFR-TH/96-35}
{\vbox{\centerline{Interactions involving D-branes}}}
\centerline{Sumit R. Das\foot{E-mail: das@theory.tifr.res.in}}
\smallskip
\centerline{\it Tata Institute of Fundamental Research}
\centerline{\it Homi Bhabha Road, Bombay 400 005, INDIA}
\smallskip
\centerline{and}
\smallskip
\centerline{Samir D. Mathur\foot{E-mail: me@ctpdown.mit.edu}}
\smallskip
\centerline{\it Center for Theoretical Physics}
\centerline{\it Massachussetts Institute of Technology}
\centerline{\it Cambridge, MA 02139, USA}
\bigskip

\medskip

\noindent
We investigate some aspects of the spectrum of D-branes and their
interactions with closed strings. As argued earlier, a collection of
many D-strings behaves, at large dilaton values, as a single 
multiply wound string. We use this result and T-duality
transformations to show that a similar phenomenon occurs for effective
strings produced by wrapping p-branes on a small (p-1)-dimensional
torus, for suitable coupling.  To understand the decay of an excited
D-string at large dilaton values, we study the decay of an elementary
string at small dilaton values. A long string, multiply wound on a
circle, with a small excitation energy is found to predominantly decay
into another string with the same winding number and an unwound closed
string (rather than two wound strings).  This decay amplitude agrees,
under duality, with the decay amplitude computed using the Born-Infeld
action for the D-string.  We compute the absorption cross section for
the D-brane model studied by Callan and Maldacena.  The absorption
cross section for the dilaton equals that for the scalars obtained by
reduction of the graviton, and both agree with the cross section
expected from a classical hole with the same charges.

\Date{July, 1996}
\def\TD{{T^{(D)}}}
\def\TS{{T^{(S)}}}

\def\tpsi{{\tilde{\psi}}}

\def\talpha{{\tilde \alpha}}
\def\tpsi{{\tilde \psi}}

\def\cA{{\cal A}}
\def\TS{{T^{(S)}}}
\def\TD{{T^{(D)}}}
\def\LS{{L^{(S)}}}
\def\LD{{L^{(D)}}}
\def\TM{{T^{(M)}}}
\def\LM{{L^{(M)}}}

\newsec{Introduction}

We have learnt over time that there exists a large class of solitonic
objects in string theory which are essential to give the theory its
dual nature. A subset of these known as D-branes
\ref\dbranes{For a recent review and references to original
literature see J. Polchinski, S. Chaudhuri and C.V. Johnson, ``{\it
Notes on D-branes}'', hep-th/9602052.}, carrying Ramond-Ramond
charges, can be studied through open strings that have Dirichlet
boundary conditions on hypersurfaces in spacetime which represent the
locations of these extended solitons. Such open strings represent the
possible excitations of the soliton, in particular the collective
modes of its low energy deformations.  The low energy field theory of
these open strings is the gauge field theory described by open
strings, dimensionally reduced to the worldvolume of the D-brane.

Thus each D-brane carries in particular a $U(1)$ gauge field on its
 surface. Witten \ref\witten{E. Witten, {\it Nucl. Phys.},{\bf B460}
 (1996) 541.} conjectured that when two D-branes approach each
 other, there is an enhancement of symmetry from $U(1) \times U(1)$ to
 $SU(2)$, due to open strings that can stretch from one brane to the
 other. More generally, with $N$ D-branes close to each other, the
 symmetry is enhanced to $U(N)$.

 This is a valid picture at weak elementary string coupling. Let the
 coupling constant of closed string theory be $g=e^\phi$.  The tension
 of D-branes is proportianal to $1/g$, so at small $g$ the D-branes
 are heavy, and can be well localised in space, with the velocities from
 quantum fluctuations being small. In
 this case one can use the approximation that the D-branes are at
 fixed locations in space, and as close to each other as we wish. The
 ends of the open string can lie on any one of the $N$ branes, and
 thus we associate a Chan-Paton factor taking $N$ values with each end
 of the open string. This gives rise to open strings describing a
 $U(N)$ symmetry group. In such calculations one usually assumes that
 the D-branes are infinitlely long along the directions which lie
 within the brane surface.

Some of the important physical applications of
D-branes require, however, an understanding of these objects at strong
coupling and wound around compact directions. 
One such application involves regarding configurations of
D-branes as black branes or black holes. So long as the branes are extremal -
corresponding to BPS states - it is possible to make exact statements
of e.g. the degeneracy of states leading to an understanding of the
Hawking-Beckenstein entropy
\ref\stromvafa{A. Strominger and C. Vafa,
hep-th/9601029}\ref\horo{ For a review of black hole entropy in string
theory and references, see G. Horowitz, gr-qc/9604051}.  However if we
are interested in the absorption of matter leading to the formation of
nonextremal states or in the decay of a non-extremal state leading to
Hawking radiation one is forced to consider the physics at
non-infinitesimal coupling.  This is because at weak coupling the
`thickness' of the soliton will be larger than the Schwarzschild
radius of the soliton, and one will be studying excitations and
de-excitations of an ordinary soliton (rather than a black hole) when
one studies interactions of closed strings with open strings that are
attached to the soliton.

However, it was shown in \ref\dasmathur{S.R. Das and S.D. Mathur,
hepth/9601152.} that at sufficiently strong coupling the nature of the
excitation spectrum for D-strings is not that suggested by the above
description of Chan-Paton factors. Consider the Type IIB theory on
$M^9 \times S^1$, with the compactified direction having length
$L$. Take a collection of D-strings wound around this compact
direction having a total RR charge equal to $n_w$.  By S-duality of
the Type IIB theory such D-strings at strong coupling should behave
exactly like a macroscopic elementary string at weak coupling. The
normalisable state of the latter is however known to be a string wound
$n_w$ times so that in particular its excitations live on a circle of
size $n_w L$. Thus the D-string with RR charge $n_w$ is represented at
strong coupling by a single string wound $n_w$ times rather than $n_w$
singly wound D-strings.  If we bring $n_w$ D-strings close to each
other to form a bound state they would join to form a single multiply
wound string such that the low energy excitations are collective modes
of a long string of length $n_w L$.

This fact becomes especially important for excited D-brane configurations
which represent nonextremal black branes 
with nonzero horizon area in the extremal
limit, e.g. D-strings bound to 5 D-branes, as considered in
\stromvafa\ and \ref\callanmalda{C.G. Callan and J. Maldacena,
hepth/9602043.}. This configuration corresponds to 
a five dimensional black hole.
The degeneracy of states with given
charges is in exact agreement with the Hawking-Beckenstein formula,
both for extremal \stromvafa\ and slightly nonextremal holes
\ref\horstom{G. Horowitz and A. Strominger, hep-th/9602051},
thus realizing the program of \ref\sussk{L. Susskind, hep-th/9309145;
J. Russo and L. Susskind, {\it Nucl. Phys.}~{\bf B437} (1995) 611;
M. Duff, M. Khuri, R. Minaisan and J. Rahmfeld, {\it Nucl. Phys.}
~{\bf B418} (1994) 195; M. Duff and J. Rahmfeld, {\it Phys. Lett.}
~{\bf 345B} (1995) 441.} and \ref\sen{A. Sen,
{\it Nucl. Phys.}~{\bf B440} (1995) 421 and {\it Mod. Phys. Lett.}
~{\bf A10} (1995) 2081; F. Larsen and F. Wilczek, hep-th/9511064}.

It was argued in \ref\susskindmalda{J. Maldacena and
L. Susskind, hepth/9604042.} that the thermodynamics of open string
states used in \callanmalda\ to obtain the extremal and nonextremal
entropies is correct for the ``fat'' black hole limit
only when one considers the branes as single
branes which are multiply wrapped around the compact direction. A
related phenomenon was found by \ref\maldacena{J. Maldacena,
hepth/9605016.} where a D-string lies within a collection of $Q_5$
parallel close by 5-D-branes.  It was argued that  the excitation spectrum 
of the D-string equals that  of a single long string with tension $1/Q_5$
 and total length $n_w Q_5$, where $n_w$ is the RR charge of the D-string.

In this paper we do the following:

\item{(a)}We investigate whether higher dimensional  branes also share
 this property of D-strings that the bound state of a  collection of 
them behaves as if it were one `long' brane instead of a collection 
of closely spaced parallel branes. By using T-dulaities
 we relate the D-string spectrum to the spectrum of  $(n_c+1)$-D-branes 
 wrapped on an $n_c$ dimensional torus. The size of this torus is small, 
so we obtain an effective string in the remaining directions, with this
 effective string turning out to behave as one long string for
a suitable range of parameters like coupling and size of wrapping torus.
 In particular  for a 
bound state of $n_w$ 5-D-branes, even at $g<<1$,
 the excitations are not those suggested by naive Chan- Paton
 factors at the ends of open strings, if the dimensions 
of the wrapping torus are less than
$g\sqrt{\alpha'}$ -- the effective string behaves as a long string 
of length $n_w$ times the length of the direction on which the
 effective string is wrapped.

\item{(b)}We wish to investigate the amplitudes for  multiply wound
 D-strings to absorb and emit massless quanta. If we had only one D-string
(RR charge  $n_w$ equal to unity) then we can study its interactions
 using open strings (attached to the D-string) 
 and closed strings (travelling througout spacetime) by an 
examination of elementary string diagrams. This is less clear 
when there are many D-strings close to each other and the  elementary
 string interaction  $g$
is strong enough to invalidate a naive description where we only 
add Chan-Paton factors to the ends of the open string. But in the 
large $g$ limit  we can make use of 
the duality with the elementary string, whose interaction cross 
sections we do know how to compute. 

 We take an initial state of the closed elementary string to be have
winding number $n_w$ around the compact direction, together with
a simple choice of oscillator excitation. Such an excited state can
decay in two possible ways. In the first the
closed string splits into two closed strings with winding numbers
 $n_{w}^{(1)}>0$ and $n_{w}^{(2)}>0$ ($n_{w}^{(1)}+n_{w}^{(2)}=n_w$)
 and no oscillator excitations; the relative 
momentum of the decay products carries
 the initial energy of excitation. 
In the second  mode we just get a string with winding
 number $n_w$ with no oscillator excitation, and the energy 
is carried away by a  massless graviton (winding number zero).
 The second process is  found to have a much larger amplitude if 
 $n_w L>> \sqrt{\alpha'}$,
and the excitation energy is small compared to the rest mass of initial 
 wound string.
\medskip

(c)\quad We use the Born-Infeld action for the D-string to
 compute the amplitude for a long wavlength vibration on a 
 D-string to decay by emitting a graviton or a dilaton. This amplitude agrees 
 under duality with that found in the corresponding process for the
 elementary string in (b) above.
\medskip

(d) \quad The results of (b) and (c) above provide some justification
for an assumption used in \ref\dasmathurtwo{S.R. Das and S.D. Mathur,
hep-th/9606185} for computing the emission from a 5-brane and a
collection of D-strings \callanmalda , which was compared with black
hole emission.  The assumption used was that the D-string excitations
behave as the excitations of one long string, and that the decay
amplitude for low energy emission is provided by the BI action for the
string.  We calculate here the cross section for absorption of low
energy scalars into the 5-brane-1-brane model, and obtain (as
expected) the same result as found through computation of emission in
\dasmathurtwo\ .  The dilaton is found to have the same absorption
cross section as the scalars coming from the Kaluza-klein reduction of
the graviton.

\newsec{Excitation spectrum of a multiply wound string.}

Let us review the argument of \dasmathur. Consider a type 
IIB elementary string, 
for the sake of definiteness. We compactify the direction 
$X^9$ on a circle of
length $L$. To lowest order in coupling, the mass of an 
elementary string state
is given by
 \eqn\oneone{\eqalign{m^2 &= (n_wL\TS+{2\pi n_p\over L})^2
+ 8\pi \TS(N_R-\delta_R)\cr
& = (n_wL\TS-{2\pi n_p\over L})^2+8\pi
\TS(N_L-\delta_L)}} 
Here $\TS = T e^{{\phi_0 \over 2}}$ is the tension of the 
elementary string. 
$n_w, n_p$
are integers giving the winding and momentum in the $X^9$ 
direction.
 $\delta_{L,R} =0,1/2$ for the Ramond and Neveu-Schwarz 
sectors
respectively.  

Let us  consider very long strings, so
that $L{\sqrt{\TS}} >> 1$. Consider the lowest excitation that 
has no net momentum: $n_p=0$. Thus $(N_R-\delta_R)=(N_L-\delta_L)=1$.
An ``unexcited'' string has
winding number $n_w$, no momentum (thus in particular $n_p=0$),
 and
$(N_R-\delta_R)=(N_L-\delta_L)=0$. The energy of the excitation 
over the energy of
this unexcited state is 
\eqn\fone{\delta m=\sqrt{(n_w\TS L)^2+8\pi \TS}-(n_w\TS L)\approx 
{4\pi\over n_w L}}

This result corresponds to the transverse vibrations of a string
 of length $n_wL$.
But this result, valid for $g\rightarrow 0$, must hold, 
by duality, also for the
D-string in the limit $g\rightarrow\infty$. Thus if we
 have a threshold bound state
of $n_w$ D-strings, at large elementary string coupling 
($g\rightarrow\infty$), then
the excitation spectrum of this state must correspond 
to that of one long string of length
$n_w L$. Indeed it was shown in \dasmathur\ that the ensemble
of such open strings with fractional momenta, but with a
total momentum which is integer (in units of $2\pi/L$)
leads to an entropy which agrees with that of the
elementary string spectrum.

This excitation spectrum is to be contrasted with the spectrum
obtained by attaching Chan-Paton factors to the ends of the open
strings that give excitations of the D-string. This would represent closely
spaced but separately  wound D-strings. The lowest energy of
such an open string is ${2\pi\over L}$, and the lowest excitation that
has vanishing total momentum is ${4\pi\over L}$ (one open string
travelling in each direction on the D-string).  The Chan-Paton factors
give this energy level a multiplicity of $n_w^2$. It is possible that
this is the correct representation  of the D-string spectrum 
at $g\rightarrow 0$; in that case  as
$g$ increases we must have a splitting of degenerate levels to a set
of levels that correspond to one long string.

\newsec{Other branes and duality}

As observed above the multiply wound D-string, at large $g$, has the
excitation spectrum of one long string rather than many singly wound
strings. It would be good to see an analogue of this for higher
branes, and in fact for various combinations of branes. Here we
provide a modest result in this direction.

We will start with a D-string wound around a compact direction, and
further assume that a certain number $n_c$ of additional directions
are compactified on circles. We choose the coupling and lengths of
compact directions such that we know that the spectrum of the D-string
is that of one long string. Now we T-dualise in the $n_c$ compact
directions, thus creating a $(n_c+1)$-D-brane from each D-string. The
energy spectrum of excitations will of course remain the same. Note
that the coupling would change under the T-duality, and in fact the 
$(n_c+1)$-D-branes obtained will have all the $n_c$ compact directions
very small, so we would have obtained an effective string, made from
wrapping $(n_c+1)$-D-branes on $n_c$ small compact directions. It is for
these effective strings that we would have established that the
excitation spectrum is that of a single long string.

Note that 

$$\TD=\TS g^{-1}$$ 
We define also the length scale
associated with these tensions 
$$\LD=\LS g^{1/2}$$ 
where $\LS$ is such
that under T-duality a circle of length $\lambda\LS$ goes to a circle
with length $\lambda^{-1}\LS$.

We start with the following fact. Take
 the elementary string, at  $g<<1$,  
wrapped on a circle of length  $L$ (direction $X^9$) 
that is order $\LS$ or larger, 
and with any other compact dimensions also having 
length $L_c$ of order  $\LS$ or larger:
$$L=A \LS, ~~~~~A>1$$
$$L_c=B\LS, ~~~~~B>1$$
Such a string with winding number $n_w$ has an excitation 
spectrum of one long string. 
 Here the restrictions on the lengths of compact directions are
imposed because if some direction becomes
sufficiently small, the one loop corrections to the mass of 
the elementary string can become large, even for small $g$.  
This is because the light states of the
string wrapping around that compact direction propagate in
higher loop string diagrams. (Such higher loop corrections 
would of course be exactly zero if $g=0$, but for the dualities 
that we are about to use we need to start with nonzero $g$.)

The S-dual of this configuration is a D-string multiply wrapped in
 the $X^9$ direction, with $X^9$ compactified on a circle of length
$$L=A \LD=A\LS g^{1/2}$$
(with $A>1$ as above).  The  other $n_c$ compact 
directions are $X^{9-n_c},\dots, X^8$.
These are on circles of length
$$L_c=B \LD=B\LS g^{1/2}$$
(with $B>1$ as above).
The dual coupling is $g_D=g^{-1}$. We take $g^D<<1$, which means $g>>1$.
Then the spectrum is that of a single long D-string,
with energy threshold
$$E_T={4\pi \over A n_w \LD}$$

Under a T-duality, the coupling changes to
$$g'=[L'/L]^{1/2}g$$
If we T-dualise all the $n_c$ directions above, then  
length of these compact directions  and the value of the coupling change 
$$L_c'=B^{-1}\LS g^{-1/2}$$
$$g'=gB^{-n_c}g^{-n_c/2}=B^{-n_c}g^{1-n_c/2}$$

Since $g>>1$, and $B>1$, the new lengths of the $n_c$ compact
directions are much smaller than the string length $\LS$. The
resulting branes describe an  effective  string extending in the $X^9$
direction, with a tension (mass per unit length) given by 
$$\TM=\TS
{g'}^{-1}({L_c'\over \LS})^{n_c} =\TS g^{-1}=\TD' B^{-n_c}
g^{-n_c/2}$$

The length of this effective string is 
$$L=A\LD=A \LD'
(g'/g)^{-1/2}=A\LD' B^{n_c/2} g^{n_c/4}=A\LM$$
  where $\LM=\LS g^{1/2}$ is the
length scale associated with the tension $\TM$.  Thus the length of
this effective string is $A>1$ times the length scale set by its own
effective tension, just like the D-string that we started with. Note that
\eqn\fone{{L_c'\over \LM}~=~B^{-1}g^{-1}~<<~1   }
so that the compactification torus for the branes is indeed much
smaller than the length scale defined by the tension of the effective
string, thus justifying the statement that we have obtained an
effective string rather than a higher dimensional brane.

As a paricular example take $n_c=4$, so that we get a 5-D-brane
wrapped on a small $T^4$, giving an effective string in the remaining
directions that is magnetically charged under the RR gauge field. (The
D-string was electrically charged under this gauge field.) The new
coupling is 
$$g'=B^{-4}g^{-1}$$ 
Since $B>1$ and $g>>1$, we have
$g'<<1$. Thus at weak elementary string coupling $g'$, suppose we take
$n_w$ 5-D-branes, and wrap 4 directions on circles of length
$L_c'=B^{-1}\LS g^{-1/2}$. Then (even though the elementary string
coupling is small) we will find that the excitations are not given by
open strings moving in an $X^9$-direction box of length $L$; instead
they are given by open strings moving in an $X^9$-direction box of
length $n_w L$.

Note that by these duality arguments we have not been able to say
 anything about the excitation spectrum of 5-D-branes with all
 dimensions of
brane large, and $g>>1$. In this latter case we do not expect
 the spectrum of a 1-dimensional object, so duality arguments
 starting from the elementary string are unlikely to access this domain.            

\newsec{Decay amplitudes for the elementary string}

Let the spacetime be $M^9\times S^1$ with $X^9$ compactified on a circle of length $L$. 
Consider an elementary string  state with  winding number $n_w^{(1)}$ around $X^9$, 
 zero momentum along $X^9$ as well as all other directions,  and an excited state of
oscillators.  There could be two possible channels for the decay
of this excited state:

(a)\quad   The initial excited state may decay into the
`ground state' with the same winding number $n_w^{(1)}$, while emitting a 
closed string
state with zero winding. 

(b) \quad  The initial state can decay into two closed strings 
each with nonzero winding number (say $n_{w}^{(2)}, n_{w}^{(3)}$,
 with $ n_{w}^{(2)}+ n_{w}^{(3)}=n_w^{(1)}$). If the initial excitation 
 had the lowest allowed energy, then these final states will
 have no oscillator excitations; the initial oscillator energy
 will be manifested as the energy of relative motion of the two final strings.

 We would like to know which of these
decay modes dominates. By duality this will tell us the dominant
decay mode of a multiply wound D-string, at $g=e^\phi >>1$.

We use the NSR formalism, and investigate the decay
of a particular class of initial excitations. The results should indicate
 the physics for an arbitrary excitation.

The fields on the closed string world sheet have the mode expansions
\ref\gsw{M. Green, J. Schwarz and E. Witten,
{\it Superstring Theory}, Vol. 1 (Cambridge University Press,
1987)}
\eqn\ttone{X^\mu_{L} = \half[x^\mu + {1\over \pi \TS}p_{L}^\mu
(\tau - \sigma) + {i \over {\sqrt{\pi \TS}}}\sum_{n \neq 0}
{\alpha^\mu_n \over n}e^{-2in(\tau-\sigma)}]}
and similarly for $X^\mu_R$ (with $\tau - \sigma \rightarrow \tau + \sigma$
and $\alpha \rightarrow {\tilde \alpha}$). For the fermion fields
we have (for the NS sector)
\eqn\tttwo{\psi^\mu_- = {1\over {\sqrt {\pi \TS}}}\sum_{r = Z + \half}
\psi^\mu_r e^{-2ir(\tau-\sigma)}}
and similarly for $\psi^\mu_+$ (with $\tau - \sigma \rightarrow \tau + \sigma$
and $b \rightarrow {\tilde b}$).

We take the initial excited state to be 
\eqn\tone{ |I> = \eta_r\tilde \eta_{r'}\epsilon_{aa'}^{(1)}{\alpha_{-N}^r
\over {\sqrt{N}}}\psi_{-1/2}^a {\talpha_{-N}^{r'}
\over {\sqrt{N}}} \tpsi_{-1/2}^{a'}  |k_{1L},k_{1R}>}
This state has $N_L = N_R = N$.
The final state is taken to be
\eqn\ttwo{|F> = \epsilon_{cc'}^{(3)}\psi_{-1/2}^c\tpsi_{-1/2}^{c'} 
|k_{3L},k_{3R}>}

The vertex operator for the state which is emitted is
\eqn\tthree{ V (u,v) = \epsilon_{bb'}^{(2)}
 e^{ik_{2L}X_L}~e^{ik_{2R}X_R}~
[\partial_u X^b + \half (k_{2L} \cdot \psi)\psi^b]~
[\partial_v X^{b'} + \half (k_{2R} \cdot \tpsi)\tpsi^{b'}]}
where $(u,v)$ are the related to the coordinates on
the cylinder by $u = \tau + \sigma$ and $v = \tau - \sigma$.

In the above the polarisations are normalised by
\eqn\normalisation{\epsilon_{ab}^{(i)}\epsilon^{(i)ab}=1, ~~~
\eta_r\eta^r=1,~~~\tilde \eta_r \tilde \eta^r=1}

To lowest order in the elementary string coupling $g$ the decay
amplitude is then given by
\eqn\tfour{ \cA = 8(\pi \TS)^2\kappa<F | V(0,0) |I>}
and may be easily evaluated to be 

\eqn\tfive{\eqalign{\cA = &8(\pi \TS)^2\kappa 
\epsilon_{aa'}^{(1)}\epsilon_{bb'}^{(2)}
\epsilon_{cc'}^{(3)}\eta_r\eta_{r'}\cr
&[{\sqrt{N}\over {\sqrt{\pi \TS}}}\delta_{ac}\delta_{br}- {1\over 2\pi
\TS} {k_{2L}^r\over 2\sqrt{\pi \TS}\sqrt{N}}(k_{1L}^b\delta_{ac}+
k_{2L}^c\delta_{ab}-k_{2L}^a\delta_{bc})]\cr 
&[{\sqrt{N}\over
{\sqrt{\pi \TS}}}\delta_{a'c'}\delta_{b'r'}- {1\over 2\pi
\TS}{k_{2R}^{r'}
\over 2\sqrt{\pi \TS}\sqrt{N}}(k_{1R}^{b'}\delta_{a'c'}+
k_{2R}^{c'}\delta_{a'b'}-k_{2R}^{a'}\delta_{b'c'})]}}

The overall coefficient in \tfour\ has been
fixed by comparing with the three graviton vertex which follows from
the Einstein action in the following way. The Einstein action for the traceless
components of the metric in the harmonic gauge becomes, upto terms
with three gravitons ($g_{ab}=\eta_{ab}+h_{ab}$)
\eqn\ttthree{S = {1 \over 8\kappa^2}[(h_{ab,c}h^{ab,c})
- (h_{ab,c}h^{ab}_{,d}h^{cd}+2h_{ab,c}h^{c,b}_d h^{ad})]} 
Consider the
amplitude for a process where a graviton $h_{12}$ with momentum $k_1$
goes into a graviton $h_{12}$ with momentum $k_3$ and a graviton
$h_{34}$ with momentum $k_2$. The tree level answer may be easily
computed from \ttthree. To do this we have to remember that we have to
use fields which are properly normalized (i.e. have the standard
kinetic energy term). In particular the off diagonal metric components
have the standard kinetic term after the rescaling $h_{12}
\rightarrow {\sqrt{2}}\kappa h_{12}$ etc. The result for this process is
\eqn\ttfour{A = {4\kappa \over 2{\sqrt{2}}}
(k_1^3k_3^4+k_1^4k_3^3)} 
This has to be compared with the string
theory answer for the three graviton vertex with polarizations
$\epsilon_i^{ab} = \epsilon_i^{ba}$ with $i = 1, \cdots 3$ and the
only nonzero components being $\epsilon_1^{12}=\epsilon_1^{21}
=\epsilon_2^{34}= \epsilon_2^{43} =\epsilon_3^{12}
=\epsilon_3^{21}={1\over \sqrt{2}}$.  The string theory answer (see
e.g. \gsw\ ) can then be seen to agree with the Einstein gravity
answer with the normalization given in \tfour.

In the following we concentrate on the case where the emitted
state has zero momentum in the string direction $X^9$. Then in the
rest frame of the initial elementary string the various momenta
are
\eqn\tsix{\eqalign{& k_{1L} = (k_1^0, {\vec 0}, n_w^{(1)}L \TS )
~~~~~~~~~~k_{1R} = (k_1^0,{\vec 0}, -n_w^{(1)}L \TS )\cr
&k_{2L} = (k_2^0, {\vec k_2}, n_w^{(2)}L \TS )
~~~~~~~~~~k_{2R} = (k_2^0, {\vec k_2}, -n_w^{(2)}L \TS )\cr
&k_{3L} = (k_3^0, {\vec k_3}, n_w^{(3)} L \TS)
~~~~~~~~~~k_{3R} = (k_3^0, {\vec k_3}, -n_w^{(3)}L \TS )}}
with $\vec k_2+\vec k_3=0$. If the emitted state is a closed string with no
 winding one
has to set $n_w^{(2)} = 0$. In \tsix\ $k_i^0$ stands for
the on-shell values
\eqn\tseven{\eqalign{& k_1^0 = {\sqrt{(\TS L n_w^{(1)})^2 + 8\pi \TS N}} \cr
&k_2^{(0)} = {\sqrt{(\TS L n_w^{(2)})^2 + {\vec k_2}^2}}\cr
&k_3^{(0)} = {\sqrt{(\TS L n_w^{(3)})^2 + {\vec k_3}^2}}}}

An interesting feature of the amplitude is that in the special case
where all the polarizations are transverse to the string direction
$X^9$, $\cA$ is independent of the values of $n_w^{(i)}$.  In
particular it does not depend on whether the emitted state is a wound
string or a graviton like state.  In the decay rate all such
differences would arise from the normalizations of the states (which
involve $k_i^0$ and hence the $n_w^{(i)}$) and phase space factors.

It is now easy to see why  for $L>>\sqrt{\alpha'}$ the decay into two wound
 strings is suppressed
relative to decay into a wound string and a massless state. The factor
in the decay rate coming from the normalizations of the states is
\eqn\teight{{\cal N}= [(2k_1^0 V)(2k_2^0 V)(2k_3^0 V)]^{-1}}
Consider the case where the transverse momenta ${\vec k_2}$ are small
compared  to $\TS L$. Then for emission of a wound state one has
$ k_i^0 \sim n_w^{(i)}L\TS$ for all $i$ so that the factor \teight\ is

\eqn\normtwo{{\cal N} \approx [8 n_w^{(1)}n_w^{(2)}n_w^{(3)}(L\TS)^3~V^3]^{-1}}
 On the other hand for the emission of a massless state one has
$n_w^{(2)}=0$ and $k_2^0 = |{\vec k_2}|$ so that the factor from
normalization is
\eqn\normthree{{\cal N} \approx [8 n_w^{(1)}n_w^{(3)}
|{\vec k_2}|(L\TS)^2~V^3]^{-1}}
 Thus emission of a state with nonzero winding is suppressed by a
factor of order $\sim |k_2|/(L\TS)$ compared to the emission of a
graviton with no winding.

At very low energies, only the first terms in \tfive\ 
contribute. These terms have the feature that the polarizations of
the initial and final states, $\epsilon^{(1)}, \epsilon^{(3)}$ do not
 affect  the polarization of the emitted state. 
In particular when
the polarization of the macroscopic string state is unchanged
by the emission we have the particularly simple answer at low
energies
\eqn\tnine{8(\pi \TS)^2{\kappa N\over \pi \TS}~\epsilon^{(2)}_{rr'}
\eta^r\tilde  \eta^{r'}}
Thus the amplitude for an excitation with left and right polarisations
 $i$, $j$, to emit the graviton $h_{12}$, is
\eqn\cfour{ 8(\pi \TS)^2{\kappa N\over \pi \TS}{1\over \sqrt{2}}  }
where we have used that $\epsilon_{12}=\epsilon_{21}=1/\sqrt{2}$.

It may be also checked that for small $k_2$ the leading contribution
at a given oscillator level comes from the type of state considered
above, rather than states of the same oscillator level obtained
by applying multiple creation operators like 
$\prod_i \alpha_{-m_i} |0>$ with $\sum_i m_i = N$. This is because
for such states with $i > 1$ there cannot be any term in the
amplitude which is independent of $k_2$.

We conclude that for $g=e^\phi >>1$ a multiply wound D-string would
preferentially decay  by emitting a massless closed string states like a
graviton rather than split into pairs of strings each with  nonzero winding.
 Thus it makes sense
to study the decay of a nonextremal D-string into gravitons and
examine to what extent this resembles Hawking radiation.

We can also compute the amplitude for the excited string to emit
a dilaton. The polarisation tensor is\ref\poloo{J. Polchinski,
{\it Nucl. Phys.},{\bf B289}
 (1987) 465.}
\eqn\done{\epsilon_{\mu\nu}~=~{1\over \sqrt{8}}[\eta_{\mu\nu}-l_\mu
k_\nu-l_\nu k_\mu]  }
where $k_\mu$ is the momentum of the dilaton, and $l_\mu$ is any null  vector
 satisfying $k_\mu l^\mu=-1$. 

Then we see that if we have an excitation of
the initial string with polarisations $i$ on each of the
left and right sides, then the amplitude
to emit a dilaton is
\eqn\cpfive{ 8(\pi \TS)^2{\kappa N\over \pi \TS}{1\over \sqrt{8}}  }
Note that this amplitude is $1/2$ times the amplitude found for the
graviton emission above.

We close this section by noting that the emission of quanta of the
 axion field $B_{\mu\nu}$ is as likely as the emission of
 gravitons. In fact the emission is to coherent superpositions of the
 graviton and the axion; the emitted quanta from the excited string
 state \tone\ are of the form ${1\over \sqrt{2}}[h_{12}+B_{12}]$. The
 vertex operator for emission of such a quantum is given by {} with
 $\epsilon_{12}=1, \epsilon_{21}=0$. (Thus
 $\epsilon_{ab}\epsilon^{ab}=1$ as before.)  The amplitude for decay
 to this mode is thus $\sqrt{2}$ higher than the decay to the graviton
 itself, and thus the probablility of decay per unit time is $2$ times
 the decay rate to the graviton.  Since the decay rate to axions
 equals that to gravitons, we find that the total decay rate obtained
 is the same whether we use the graviton and axion as our fields or if
 we use ${1\over \sqrt{2}}[h_{\mu\nu}\pm B_{\mu\nu}]$ as our fields.

\newsec{D-String amplitudes}

We now compare the elementary string amplitude derived in the previous
section with amplitude for emission of a massless closed string state
from an excited D-string at low energies. The amplitude for such
a process has been computed in \ref\hashikleba{A. Hashimoto and I. 
Klebanov,
hepth/9604065}. For our purposes it is most efficient to use the
Born-Infeld action to derive the result.

The DBI action which describe the low energy dynamics of a D-string
may be written in terms of the coordinates of the D-string
$X^\mu(\xi^m)$ (where $\mu$ runs over all the $10$ indices whereas
$\xi^m$ are parameters on the D-string worldsheet) and the gauge
fields on the D-string worldsheet $A^m (\xi^m)$ as follows
\ref\broninfeld{See e.g. Ref [1]; A. Tseytlin, hep-th/9602064.}
\eqn\thronep{S_{BI} = T \int d^2\xi~e^{-\phi(X)}
{\sqrt{ {\rm det}[G^{(S)}_{mn}(X) +B_{mn}(X)+ 2\pi\alpha' F_{mn}]}}}
where $F_{mn}$ denotes the gauge field strength on the D-string worldsheet
and $G^{(S)}_{mn}$  and $B_{mn}$ are given by 
\eqn\thrtwo{G^{(S)}_{mn} = G^{(S)}_{\mu\nu}(X)\partial_m X^\mu 
\partial_n X^\nu
~~~~~~~~~~~B_{mn} = B_{\mu\nu}(X)\partial_m X^\mu \partial_n X^\nu}
Here $G^{(S)}_{\mu\nu}$ is the target space metric in the string
frame, $B_{\mu\nu}$ is the antisymmetric tensor field, and $T$ is a
tension related to the tension $\TS$ of the fundamental string through
$T=\TS e^{-\phi/2}$.

We will concentrate on the coupling of the D-string to gravitons and
the dilaton, and so ignore the $B$ field and the field strength $F$
for the following calculation.  As mentioned at the end of the
previous section, the $B$ field couples to the D-string as efficiently
as the gravitons, but we may separate gravity from the $B$ field for
convenience, which is what we will do below. We also shift to the
Einstein metric in the following, given by
$G_{\mu\nu}=e^{-\phi/2}G^{(S)}_{\mu\nu}$. Then the DBI action may be
written as
\eqn\throne{S_{BI} = T \int d^2\xi~e^{-\phi(X)/2}
{\sqrt{ {\rm det}[G_{mn}(X)]}}}

The bulk action is

\eqn\cone{S_{\rm bulk}~=~{1\over 2\kappa^2}\int d^{10} x
 {\sqrt{G}}[R~-~{1\over 2}\partial \phi \partial \phi~+~
\dots..]}
where we write only the terms that we shall need.

We will work in the static gauge which means
\eqn\thrthree{X^0 = \xi^0~~~~~~~~~~~X^9 = \xi^1}
The D-string worldsheet is then the $X^0, X^9$ plane. 
In this gauge the massless open string
fields which denote the low energy excitations of the brane are
the transverse coordinates $X^i(\xi^0,\xi^1),~~i = 1,\cdots 8$. 

In the following we will set the gauge field and the RR field to be zero.
The lowest order interaction between the metric fluctuations around
flat space and the open string modes is obtained by expanding the
metric as $G_{\mu\nu} = \eta_{\mu\nu} +  h_{\mu\nu} (X)$,
expanding the transverse coordinates $X^i (\xi)$ around the brane
position $X^i = 0$ and treating $h_{\mu\nu}$ and $X^i$ to be
small.  One then has
\eqn\thrfour{\eqalign{g_{mn}& = 
[\eta_{mn}]~+~[h_{mn}]~+~[X^i_{,m}X^j_{,n}\eta_{ij}+
X^i_{,m}h_{i n}+X^j_{,n}h_{m j}] ~+~[X^i_{,m}X^j_{,n}h_{ij}]\cr
& \equiv [\eta_{mn}]~+~g_{mn}^{(1)}~+~g_{mn}^{(2)}~+~g_{mn}^{(3)}}}
The above relation is exact, but we have grouped terms according
to the order of smallness, assuming that in later use we will treat
$X^i, h_{ij}$ as being small. We have
\eqn\thrfive{\det(G)~=~ - \det [1+C]}
where
\eqn\thrsix{C=\eta^{-1}[g_{mn}^{(1)}~+~g_{mn}^{(2)}~+~g_{mn}^{(3)}]}
Note that for a $2\times 2$ matrix  $C$,
\eqn\thrseven{\det[1+C]=1+ tr C+\det C}
Then upto terms involving two open string fields we have
\eqn\threight{\eqalign{& {1\over 2}[\delta_{ij} +  h_{ij} - 
{1\over 2} h^\alpha_\alpha \delta_{ij}]
\partial_\alpha X^i \partial^\alpha X^j\cr
& + {1\over 2}
 (\partial_k h^\alpha_\alpha)X^k +
 {1\over 4}(\partial_k \partial_i
h^\alpha_\alpha) X^i X^k + 
 h_{i\alpha}\partial^\alpha X^i +
 (\partial_j h_{i\alpha})X^j \partial^\alpha X^i\cr
&-{1\over 2}
 \delta_{ij} (h_{00}\partial_1 X^i \partial_1 X^j 
+ h_{11} \partial_0
X^i \partial_0 X^j - 2 h_{10}\partial_0 X^i \partial_1 X^j)}}
For purely transverse gravitons, i.e. only $h_{ij} \neq 0$ this
simplifies to
\eqn\thrnine{{1\over 2}(\delta_{ij} +  h_{ij})\partial_\alpha X^i
\partial^\alpha X^j}

Consider a D-string which is excited above the BPS state by addition
of a pair of open string states with momenta (on the worldsheet)
$(p_0,p_1)$ and $(q_0,q_1)$ respectively.  These open strings are to be
 identified with quanta of the variable $X^i$ in the above BI action.
 Note that the field with a standard kinetic term is not $X^i$ but
\eqn\ctwo{\tilde X^i~=~\sqrt {T} X^i}
The polarizations of the
open string states are chosen to be transverse. The decay of this
state into the extremal state is given by the process of annihilation
of this pair into a closed string state, like a graviton. For a
graviton which is also transversely polarized to the D-string 
with momentum $(k_0, k_1, {\vec k})$ (where ${\vec k}$ denotes the
momentum in the transverse direction), the
leading term for this amplitude for low graviton energies can be
read off from \thrnine\ as
\eqn\thrten{ \cA_D = \lambda_{(1)}^i \lambda_{(2)}^j \epsilon_{ij}
p \cdot q}
where the polarisations are normalised as
\eqn\cfive{\epsilon_{ij}
\epsilon^{ij}=1, ~~~~~~\lambda_{(1)i} \lambda_{(1)}^i =1,
 ~~~~\lambda_{(2)i} \lambda_{(2)}^i =1}
When the outgoing graviton does not have any momentum along the
string direction one has
\eqn\threleven{p \cdot q = p^0q^0 - p^1q^1 = 2 |p_1|^2}
where we have used momentum conservation in the string direction
and the masslessness of the modes. 

Writing this another way, the amplitude per unit time for a pair of
open strings with equal and opposite momenta to collide and emit a
graviton is
\eqn\cthree{R_h~= {\sqrt{2}}\kappa (2 |p_1|^2)
{1\over \sqrt {2|p_1|}}{1\over \sqrt{L}}
{1\over \sqrt {2|p_1|}}{1\over \sqrt{L}}L{1\over 
\sqrt{2\omega_h}}{1\over \sqrt{V}}~=~\sqrt{2}\kappa |p_1|  
{1\over \sqrt{2\omega_h}}{1\over \sqrt{V}} }    
Here the first two factors come from the fact that the field with standard
kinetic term corresponding to say the graviton $h_{12}$ is
$(2\kappa){1\over \sqrt{2}}h_{12}$. 
The last factors are from the normalisations
 present in the fields and the volume of the interaction region $L$.
 ($L$ is the total length of the D-string.)

Let us also compute the amplitude for the open strings to collide and
emit a dilaton quantum. From the action \throne , the relevant
contribution is
\eqn\cfour{S_{BI}~\rightarrow~-{\phi\over 2}~ T~{1\over 2}~\delta_{ij}
\partial_\alpha X^i\partial^\alpha X^j   }

>From the bulk action \cone\ we note that the field corresponding to the
 dilaton with correctly normalised kinetic term is 
\eqn\csix{\tilde \phi~=~{\phi\over \sqrt{2}\kappa}  }
Then we find that the amplitude for two open strings (both with the
 same polarisation but with equal and oposite momenta) to collide and
 emit a dilaton is

\eqn\cseven{\eqalign{R_d~&=~ (\sqrt{2}\kappa) {1\over 4}2 
(2|p_1|^2){1\over \sqrt {2|p_1|}}
{1\over \sqrt{L}}
{1\over \sqrt {2|p_1|}}{1\over \sqrt{L}}L
{1\over \sqrt{2\omega_d}}{1\over \sqrt{V}}\cr
& =~{1\over \sqrt{2}}\kappa |p_1|  {1\over \sqrt{2\omega_d}}{1\over \sqrt{V}}}}
In \cseven\ the factor ${\sqrt{2}}\kappa$ comes from \csix\ and the factor
$1/4$ comes from the coefficient in the BI action in \cfour.
There is an additional factor of $2$ since we are considering 
the emission of diagonal
elements of $h_{ij}$ which arise from two
open string states of the same polarization. Then
each open string annihilation operator 
in the BI action term $h_{ii}\partial X^i \partial X^i$
can kill either of the initial open strings states.

Note that $R_d=R_h/2$.

\newsec{Comparison of elementary and D-string S-matrices}

We now compare the S-matrix for the decay of an excited elementary
string into a massless graviton with the S-matrix of the decay of an
excited D-string into the same polarization state.  For definiteness
we will consider the polarization state $h_{12}$.

First consider the elementary string. From the results of section 4
 (equations \tfour , \tfive\ ) we get that for $n_w L>>\sqrt{\alpha'}$
 and small excitation number $N$ (which together imply that
 $k_2^0<<\TS^{1/2}$) the amplitude per unit time to decay to the
 graviton is

\eqn\lone{\eqalign{A'_E & \approx
({2 \over 2{\sqrt{2}}})(8 (\pi \TS)^2)~
{\kappa \over \pi \TS}~N~ V [(2\TS L n_w V)^2 (2 k_2^0 V)]^{-1/2}\cr
&= {2\pi N \over n_w L}{{\sqrt{2}}\kappa \over {\sqrt{2k_2^0 V}}}}}
The origin of the various factors has been explained in section 4.  The
states are normalized in the nine dimensional
spatial volume $V$. The overall V comes from the momentum conserving
delta function after setting the initial and final momenta to be equal.

The D-string decay amplitude per unit time is similarly obtained
from \threleven\ as
\eqn\ltwo{A_D = {\sqrt{2}}
\kappa (2 |p_1|^2) L [(2|p_1|L)(2|p_1|L)(2 k_2^0 V)]^{-1/2}
= {{\sqrt{2}}\kappa |p_1| \over {\sqrt{(2 k_2^0)(V)}}}}
where the origin of the various terms is exactly the same as in
\lone\ with the difference that the open string states are normalized
on the D-string rather than in the entire space. The factor of
${\sqrt{2}}$ comes from the fact that the field with standard kinetic term
is $\sqrt{2}\kappa h_{12}$.

The D-string is believed to be dual to the elementary string. If this
is true these two amplitudes \lone\ and \ltwo\ must be equal. Under
the duality transformation the oscillator states of the elementary
string become the momentum states of the open strings on the D-string,
with the oscillator number being identified with the quantized open
string momenta. In fact the D-string answer \ltwo\ is in exact
agreement with the elementary string answer \lone\ with the
identification
\eqn\lfive{ |p_1| = {2 \pi N \over n_w L}}

In a similar way we verify that the amplitude for an excited elementary string
 to emit a dilaton equals the amplitude  for a D-string to emit a dilaton,
 using equations \cpfive\  and \cseven . 

These results show again that the emission predicted by the BI action
  for a D-string with a RR charge $n_w$ equals that for an elementary
  string at $g<<1$ that is multiply wound $n_w$ times.

\newsec{The Absorption cross-section}

In \stromvafa\ 
a model was given using D-branes which, at strong coupling, 
would correspond to an extremal black hole 
in 5 dimensions. This black hole carries three nonzero charges, 
and so has nonzero horizon area. 
The entropy of the D-brane system agreed with the 
Bekenstein entropy given by this area. In 
\callanmalda\ it was shown that a slightly 
nonextremal configuration of these branes 
radiates with the temperature expected from black 
hole thermodynamics, and moreover the 
emission rate is proportional to the 
horizon area implied by the charges of the near-extremal hole. 
In \dasmathurtwo\
it was shown that the emission of 
low energy scalar quanta (obtained by
dimensional reduction of the 10-d graviton) from the 
slightly non-extremal configuration of branes 
agreed exactly with the radiation expected from the corresponding black hole. 
(For some other results on nonextremal branes see
\ref\many{S.S. Gubser, I. Klebanov
and A.W. Peet, hep-th/9602135; I. Klebanov and A. Tseytlin,
hep-th/9604089;M.
Cvetic and D. Youm, hepth/9603147, hep-th/9605051, hep-th/9606033;
G. Horowitz, D. Lowe and J. Maldacena, hep-th/9603195; M.Cvetic and
A. Tseytlin, hep-th/9606033;
E. Halyo, A. Rajaraman and L. Susskind, hep-th 9605112.}.)

Below we compute the cross section for this collection of branes to
absorb (a) scalars derived from a Kaluza-Klein reduction of the
graviton and (b) the dilaton.  These two results (a) and (b) will be
found to agree. This agreement is essential if they are to be compared
to absorption by a black hole, since a black hole absorbs all
uncharged scalars with the same cross-section. As expected from
general arguments of detailed balance, this cross section agrees with
that computed from an analysis of emission in \dasmathurtwo .

The absorbing system is the D-string, with winding number $n_w$ 
 around $X^9$ which is compactified on a circle of length $L$. We assume,
 following \ref\hormalstr{G. Horowitz, J. Maldacena and
A. Strominger, hep-th/9603109.}, that the D-string is constrained to move
 in only four out of its eight  
transverse  directions by a 5-D-brane, to which it is bound. 
There is a thermal distribution of
momentum modes on the D-string (say left moving), 
with total momentum $2\pi N/L$.

A D-string of length $L$ may be considered as a system with some discrete
energy levels with spacing $\Delta E$ which is independent of $E$. 
Consider an inital state at $t = 0$
where the D-brane system is in its BPS
ground state and a closed string state of energy $k_0$ is incident on it.
Let the amplitude to excite the D-string to  any one of the excited levels
per unit time be $R$.
(For $t$ large, only the levels in a narrow band will contribute,
and in this band we can use the same $R$ for each level.)
 Then the amplitude that the system in an
excited state with energy $E_n$ at a given time $t$ is given by
\eqn\tenone{ A(t) = R e^{-iE_nt}\int_0^t dt' e^{i(E_n - k_0)t'}
= Re^{-{i\over 2}(E_n+k_0)t}[{2 \sin [(E_n-k_0)t/2] \over (E_n-k_0)}]}
The total number of quanta absorbed in time $t$ is thus given by 
\eqn\tentwo{P(t) = \sum_n |R|^2 [{2 \sin [(E_n-k_0)t/2 ]\over (E_n-k_0)}]^2
\rho (k_0)}
where $\rho (k_0)$ denotes the occupation probablity of the graviton
in the initial state with energy $k_0$.
For large length of the D-string $L$ we can replace the sum by 
an integral
\eqn\tenthree{\sum_n \rightarrow \int {dE \over \Delta E}}
in which case the rate of absorption $ {\cal R_A} = P(t)/t$ evaluates to
\eqn\tenfour{{\cal  R_A }(t) = {2\pi |R|^2 \over \Delta E} \rho( k_0)}

For our case of the D-string on the 5-brane, 
\eqn\aone{\Delta E~=~{4\pi\over n_w L} }
Note that because we have the spectrum of one long string
of length $n_w L$ rather than $n_w$ strings of length $L$, we have 
closely spaced levels for $n_w$ large, and thus the approximation
\tenthree\ is improved. It is possible that interactions
further smooth out the discrete level separation \aone\ towards
a continuum, but we shall not investigate this issue here.

Consider the absorption of a quantum of the graviton $h_{12}$, with no
momentum or winding along the compact directions. There are two open
string states that can be created on the D-string in absorbing this
graviton. We can have the string with poilarisation $1$ travelling
left on the D-string and the open string with polarisation $2$
travelling right, or we can have the polarisations the other way
round. This means that there are two series of closely spaced levels
that will do the absorption, and so the final rate of absorption
computed from \tenfour\ will have to be doubled.

>From \thrten\ we find for the amplitude per unit time for the graviton
to create any one of these two possible open sring configurations to
be
\eqn\atwo{R=\sqrt{2} \kappa |p_1| {1\over \sqrt{2k^0_2 }}
{1\over \sqrt{L}} {1\over \sqrt{V_c}} {1\over \sqrt{V_T}} 
\rho_L^{(1/2)}(|p_1|) }
where we have separated the term ${1\over \sqrt{V}}$ into
contributions from the string direction $X^9$, the remaining four
compact directions (denoted by the subscript $c$) and the transverse
noncompact patial directions (denoted by the subscript $T$). We have
also included the term
\eqn\afour{[\rho_L(|p_1|)]^{1/2}=[{T_L\over |p_1|}]^{(1/2)}, ~~~
 T_L={S_L\over \pi n_w L} } 
which gives the Bose enhancement factor
 due to the population of left moving open string states on the
 D-string \callanmalda .  Here $S_L$ is the entropy of the extremal
 configuration, given by the count of the possible ways to distribute
 the $N$ quanta of momentum among different left moving vibrations of
 the D-string:
\eqn\eone{S_L~=~2\pi\sqrt{n_w N} }
and equals the Bekenstein entropy of the black hole with the same
 charges as the D-brane configuration.\foot{For a derivation of the
 results of \callanmalda\ in the notation used here, see \dasmathurtwo
 .}

The absorption cross section is given by
\eqn\athree{\sigma~=~2{\cal R_A}/{\cal F}}
where ${\cal F}=\rho(k_0)V_T^{-1}$ is the flux, and the factor
of $2$ was explained before eq. \atwo. 

Note that
\eqn\afive{{\kappa^2\over LV_c}~=~8\pi G_N^5}
and that for the given choice of momenta
\eqn\gone{k_0~=~2|p_1|  }
Then we find
\eqn\asix{\sigma~=~A}
where $A=8\pi G_N^5\sqrt{n_w N}$ is the area of the extremal black
hole with one 5-D-brane, $n_w$ windings of the 1-D-brane, and momentum
charge $N$.  This result agrees, as expected, with the calculation of
\dasmathurtwo\ where the cross section was computed from the emission
of quanta from the slightly nonextremal configuration of branes, and
the result \asix\ was shown to agree with the classical cross section
for absorption of scalar quanta.

Now consider the absorption of the dilaton. As shown in sections 4,5,
 the amplitude per unit time for two open strings of the same
 polarisation to collide and emit a dilaton is $1/2$ times the
 amplitude for open strings of polarisation $1$ and $2$ to collide and
 emit the 5-dimensional scalar given by $h_{12}$. The rate of emission
 for the dilaton is thus $n/4$ times the rate for emission of $h_{12}
 $ quanta, where $n$ is the number of polarisations allowed for the
 open strings. Since we have $n=4$, the emission rate for the dilaton
 equals that for the other scalars, and repeating the above
 calculation shows that the absorption cross section will be the same
 for the two cases as well.

\newsec{Discussion}

We were interested in examining the absorption cross section for
5-dimensional scalars in the D-brane configuration discussed in
\stromvafa \callanmalda .  This configuration would give an extremal
black hole with nonzero horizon area, if the charges and the coupling
were appropriately large.

It is not clear how to access the region of parameter space that
 corresponds to the black hole through simple D-brane
 calculations. The classical cross section for absorption of waves
 into the black hole is of course computable, in particular for low
 energy waves one can follow the methods of \ref\page{D. Page, {\it
 Phys. Rev.}~{\bf D13}(1976) 198.}  or \ref\unruh{ W.G. Unruh, {\it
 Phys. Rev.}~{\bf D14} (1976) 3251.}. Such a calculation was done for
 the extremal black hole under consideration in \dasmathurtwo , and
 for a nonextremal version in \ref\spenta{A. Dhar, G. Mandal and
 S.R. Wadia, hep-th/9605234.}. As in the 3+1 dimensional Schwarzschild
 case, the cross section equals the area $A$ of the horizon.

With D-branes, what we have computed is the absorption for the case
 when we have one 5-D-brane, a given number of 1-D-branes, and
 momentum on the 1-D-branes. For $g=e^\phi<<1$, and long wavelength
 oscillations, one believes that the BI action for the D-string should
 be a good description.  We have shown that for $g>>1$, the results
 given by the BI action are still obtained. This was done using that
 fact that the D-string at $g>>1$ behaves like an elementary string at
 $g<<1$, and in the latter case we know how to compute the deacy rates
 again.  In particular in the elementary string case we know how to
 handle the issue of the decay of a multiply wound string, and the
 result agrees with using the BI action for a single long string
 rather than a collection of closely spaced individually wrapped
 strings.  This is useful because from the description of the D-brane
 excitations as open strings with ends on the D-brane, we do not quite
 know how to handle the oscillations of bound states of several
 parallel branes.  A naive placing of Chan-Paton factors at the ends
 of the open string is in fact not correct at least for large $g$,
 where duality with the elementary string gives the excitation
 threshold of a single long string.

Note that just because we know how to handle the D-string at both
 large and small $g$ does not mean that we can access the black hole
 limit.  The latter may imply values for the charges $n_w$, $N$, and
 the coupling $g$ that do not permit using the lowest order
 perturbation theory that we have done here to study the interactions
 of scalars with the D-branes.  (In particular, a D-string at large
 $g$ is just like an elementary string at small $g$, so we do not
 access the black hole limit by just taking a D-string and going to
 large $g$.)  What is interesting is that the low energy absorption
 cross section nevertheless agrees between the black hole and the
 D-brane cases. In particular the fact that only four directions of
 oscillations are allowed in the D-brane model was essential in
 getting the dilaton absorption cross section to agree with the cross
 section of the other scalars, this agreement being a basic
 requirement of black hole theormodynamics. (In the D-brane case this
 might be a consequence of the supermultiplet structure in the D-brane
 model.)  Another interesting feature is that in the 3+1 dimensional
 hole, the absorption cross section for spin-$0$ and spin-$1/2$ quanta
 are proportional to the area $A$ for low energy quanta, while the
 cross sections for higher spins vanish at low energy, being
 multiplied by higher powers of $A\omega^2$ \page .  In the D-brane
 model we get emission of scalars from the collision of two bosonic
 open string states on the D-string, and the emission of spin-$1/2$
 quanta from the collision of a left moving bosonic string and a right
 moving fermionic string.  It is important to have the left moving
 string to be bosonic, so that we get the bose enhancement factor
 \afour\ without which the cross section vanishes in the classical
 limit. We find again that only spin-$0$ and spin-$1/2$ quanta have
 nonvanishing cross sections at low energy. 

It is interesting to 
note that the computation of absorption from the classical black hole
geometry is a calculation involving only the five noncompact
directions, with the wavefunction having no nontrivial dependence in
the compactified directions. The D-brane calculation, by contrast,
involves converting the energy of the incoming graviton (which again
has a wavefunction with nontrivial dependence only in the five
noncompact directions) into a pair of open strings that have momentum
in the {\it compact} directions.  Thus these quite different
mechanisms have lead, in the present low energy calculation, to the
same cross section.  It is possible that the agreement between the
black hole and the D-brane cases might be a consequence of some
combination of the following: the supersymmetry of the extremal
configuration, the near extremality of the absorption/emission
process, and the low energy of the quanta considered. Investigation of
these issues in in progress.

\newsec{Acknowledgements} 

We would like to thank L. Susskind and B. Zweibach for discussions.
S.R.D. would like to thank the Theoretical High Energy Physics Group
of Brown University and the Center for Theoretical Physics, M.I.T. for
hospitality. S.D.M. is supported in part by D.O.E. cooperative agreement
DE-FC02-94ER40818.

\listrefs
\end